\documentclass[prl,aps,ams,twocolumn,epsfig]{revtex4-1}
\usepackage{epsfig}
\usepackage{amsmath}
\usepackage{amssymb}
\usepackage{bm}
\usepackage{dcolumn}% Align table columns on decimal point
\newcommand{\be}{\begin{equation} }
\newcommand{\ee}{\end{equation} }
\newcommand{\ba}{\begin{eqnarray} }
\newcommand{\ea}{\end{eqnarray} }

\newcommand{\bit}{\begin{itemize}}
\newcommand{\eit}{\end{itemize}}

\newcommand{\mac}{\mathcal}

\newcommand{\bv}{\left( \begin{array} }
\newcommand{\ev}{\end{array} \right ) }

\def\om{\mbox{\boldmath$\Omega$}}

\newcommand{\ch}{Chain-Mail }

\begin{document}

\title{Wilson Line Picture of Levin-Wen Partition Functions}

\author{F. J. Burnell$^{1,2}$ and Steven H. Simon$^2$ \\
$^1$All Souls College, Oxford, UK \\ ${}^2$Rudolf Peierls Centre for Theoretical Physics, Oxford, OX1 3NP, UK}

\begin{abstract}
Lattice Hamiltonians (for example, Phys.~Rev.~B 71, 045110 (2005)) can
be constructed that have a low energy description which is a doubled
Chern-Simons theory --- two independent opposite chirality topological
sectors.  We show that the partition function of these theories is an
expectation of Wilson loops that form a link in 2+1 dimensional
spacetime known in the mathematical literature as Chain-Mail.  This
geometric construction
establishes a concrete connection between the lattice models
and continuum Chern-Simons theories, allowing us to use well-established results
about the latter to obtain
 a physical interpretation of the lattice
model Hilbert space and Hamiltonian, its topological invariance,
exactness under coarse-graining, and how two opposite chirality
sectors of the doubled theory arise.
These features of the lattice models can thus be situated in the broader context
of topological invariants obtained from Chern-Simons theories.
\end{abstract}
\date{\today, 2010}
%\pacs{}
\maketitle

Topological Quantum Field Theories (TQFTs), including the well known example of Chern-Simons (CS) gauge theory, can be roughly defined as field theories where amplitudes depend not on distances, momenta, or energies, but only on topologies  -- such as the topology of the spacetime manifold or the topology of the knotting of world lines of particles\cite{Witten,RevModPhys,Kitaev}.  Over the past few years there has been increasing interest in realizing TQFTs in condensed matter systems, in a large part driven by the possibility of using them for quantum computation\cite{RevModPhys,Kitaev}.   A central question has been to determine which physical systems (even toy models) are described by a TQFT at low energy and long length scale.    A major step in this direction has been taken by Levin and Wen\cite{LevinWen} (LW), {\it who describe a construction which can produce lattice models whose  excitations correspond to those of a ``doubled" Chern-Simons (CS) theory}.  
%description is a so-called ``quantum double'' TQFT 
(See also \onlinecite{Kitaev}).   In the current paper we will  {\it establish a direct relationship between these models and a continuum Chern-Simons theory.}

The LW lattice models\cite{LevinWen} are constructed as follows. One chooses a 2d lattice, and defines the states of the Hilbert space to be all possible assignments of labels ${\bf i}_k \in \{ {\bf 1}, \ldots, {\bf r} \}$ to each edge of the lattice, where two such states are defined to be orthogonal if at least one edge label differs\cite{channels}. The LW Hamiltonian is then constructed from operators which act on these labels according to a set of rules (called fusion rules) dictated by the mathematical structure of the constituent chiral CS theory.

An obvious question that arises from such a description is what, physically, do these labels and operators represent?    In some well known cases, such as the Toric Code, there is a physical correspondence to a discrete lattice gauge theory\cite{LevinWen,Kitaev} thus giving an interpretation of the labels (although  these cases are not doubled CS theories).  Ref.~\onlinecite{Trebst} instead views edge labels as quantum numbers of anyonic liquids on closed surfaces,
which thereby constitute a  quantum double model. Here we present another approach to this question: we will show how edge labels in these theories, as well as the Hamiltonian that acts on these labels, can be interpreted as Wilson lines of the constituent CS gauge theory\cite{endnote1}.
This (chiral) CS theory is defined on the space-time of the lattice model;  however, it turns out to be equivalent to a CS theory that lives in two connected copies of this spacetime in which the gauge field has opposite chirality, thus giving rise to the doubled CS theory expected for the LW models which we consider.

Our approach leads to several insights beyond a physical interpretation of the states and operators in the Hilbert space.  First, it establishes a concrete connection between the partition functions of these models and a link invariant (the \ch invariant, introduced by Roberts\cite{Roberts}), which is known to be equivalent to the Turaev-Viro state sum invariant of 3-manifolds.  
%Hence we establish a firm connection between these lattice models and the mathematically studied topological invariants which their partition functions describe.  
{\it Numerous techniques for calculating these invariants exist\cite{Wang}; the correspondence which we establish here therefore allows for a straightforward deduction of the partition function, even in the presence of quasi-particles.}  Second, by studying deformations which do not change the link invariant (the expectation of the Wilson loop operators), one can understand in simple graphical terms which deformations to the lattice structure (such as coarse-graining) leave the partition function unchanged, and hence do not affect the physical dynamics of the lattice model.  {\it This gives an alternative perspective on coarse-graining methods known to be exact for these models\cite{LevinNaveWenGu}.}  Similarly, in this language it is immediately apparent how to construct analogous models on any lattice geometry.  Finally, by studying the quasi-particle excitations of these models in the continuum theory, we arrive at an elegant picture of their statistics: the particles are divided into two separate sectors, one associated with each (opposite-chirality) copy of the spacetime in the continuum description.  Interestingly, these quasi-particles describe possible modifications to the \ch link invariant, some of which have not been previously discussed in the mathematical literature.
A more detailed and general discussion of these results will be given in a forthcoming work\cite{Us}.

The correspondence we make hinges on the fact that the Wilson lines in a CS theory combine according to fusion rules which have precisely the mathematical fusion structure required for constructing the LW Hamiltonian.
A Wilson line in the CS theory has the form  $W_{\bf i}(C)={\rm tr}_{\bf i}[ P \exp \int_C a(\vec x)] $ where $a$ is the gauge field, $C$ is a closed curve, and $P$ indicates path ordering.  Each Wilson line carries a  label (or  ``quantum number")
${\bf i}\in \{{\bf 1},\ldots, {\bf r}\}$, denoting the representation of the quantum group ${\cal G}$ in which the trace is taken.  (${\cal G}$ is closely related to the gauge group of the classical Chern-Simons theory, but has only finitely many representations, as is appropriate for quantum Chern-Simons theory).  We will always denote the trivial (or ``vacuum") representation as  {$\bf 1$} (which can be thought of as the absence of a Wilson line, or zero flux).  Two Wilson lines are fused by combining the gauge group representations that they carry --- the result being expressed as a superposition of single Wilson lines, whose components are representations drawn from the set of possible outcomes of this particular fusion process.  This leads to a set of ``allowed" trivalent vertices of Wilson lines, which are consistent with the fusion rules\cite{Witten,Kitaev,RevModPhys,Wang}.   For example, if ${\cal G} = SU(2)$,  the labels correspond to the total spin of the representation.  Fusion is then a generalization of the fact that two spin-$\frac{1}{2}$ representations can be combined to give either singlet or triplet states.
Unlike the familiar example of adding angular momenta, in CS theory the number ${\bf r}$  of allowed representations is finite.

Since CS theory is a TQFT, the expectation of any product of Wilson loop operators gives a result that depends only on the topology of the path -- i.e., the strings can be deformed continuously and the result remains unchanged as long as the strings do not cut through each other.  Thus the expectation can be thought of as a ``link invariant" of the input labeled link of Wilson lines\cite{Witten}.

The innate fusion structure of the CS theory allows us to construct a LW Hamiltonian, and understand the ground state structure of these models, in the language of Wilson lines.  Specifically, the edge labels of the LW model are taken to be the allowed representations $\{{\bf 1},\ldots, {\bf r}\}$ of the Lie algebra of the CS gauge group, and a fusion structure for these labels is inherited from the representation theory of this group.
 The LW Hamiltonian, most simply expressed on the honeycomb lattice, has the form $H= E_G \left( \sum_{\mbox{\tiny vertices } m} (1-V_m) +\sum_{\mbox{\tiny plaquettes } n} (1-P_n) \right ) $   where $E_G$ is the gap energy scale and all of the $V$'s and $P$'s are mutually commuting projectors.    The ground state
 thus satisfies the constraints $P_n=1$ for each plaquette and  $V_m=1$ at each vertex.  The vertex constraint $V_m=1$
 forces the three labels incident on vertex $m$ to fuse to $\bf 1$, meaning that this trivalent vertex is allowed by the fusion rules.
 It is thus tempting to view this ground state as a network of Wilson lines (a ``string-net" in the language of Ref.~\onlinecite{LevinWen})--- which is
 what we shall do, although we will eventually work in 2+1d rather than just in 2d.

\begin{figure}
\begin{center}
\includegraphics[width=0.8\columnwidth]{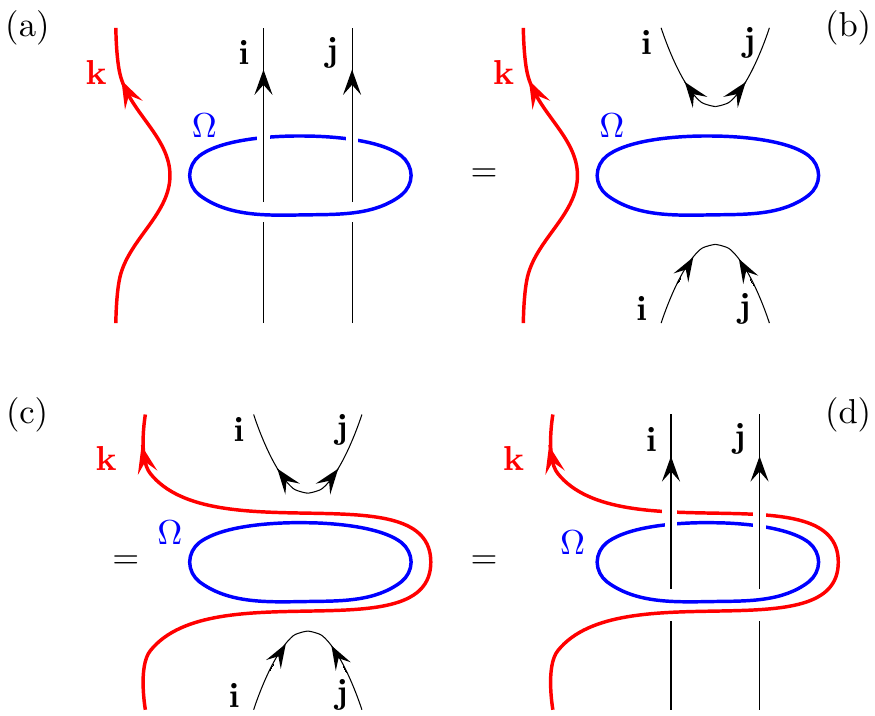}
\end{center}
\caption{(color online) The $\om$ loop as a projector.  All four diagrams of Wilson lines give the same expectation (i.e., the same link invariant).  Here $\bf i$ and $\bf j$ must fuse to $\bf 1$ (so $\bf i$ is the conjugate, or antiparticle, of $\bf j$). More generally if multiple Wilson lines thread an $\om$ loop they must fuse to $\bf 1$ as a group.
%Going from (a) to (b) or (d) to (c) are examples of projection to trivial flux -- only the vacuum particle $\bf 1$ can go through the $\om$ loop.
Going from (a) to (d) is an example of handleslide.  }\label{KillingHandle}
\end{figure}

The  projector $P_n$ is a dynamical term that flips labels on the edges of plaquette $n$, such that the ground state is a superposition of all possible string nets.   In the language of Wilson lines, the action of a plaquette term is equivalent to running a loop carrying a particular superposition of Wilson lines (which we call $\om$) around the perimeter of a hexagonal plaquette and fusing this loop into the Wilson lines running along the six edges.
This action corresponds to
combining the gauge flux of the new Wilson line with that of the existing state without destroying the string-net structure required by the $V_m$'s.  Crucially, we will also be able to describe the action of the vertex projectors $V_m$ in the language of same object, $\om$.
 To be precise, $\om$ is the superposition of quantum numbers
 $\om = \sum_{\bf i={\bf 1}}^{\bf r}  (d_{\bf i}/{\cal D}^2)  {\bf i}$ where $d_{\bf i}$ is the so-called quantum dimension of $\bf i$ (the expectation of a single unknotted Wilson line loop in representation $\bf i$) and ${\cal D}^2 = \sum_{\bf i=1}^{\bf r} d_{\bf i}^2$.  This superposition is chosen so that $\om$ loops act as projectors onto trivial flux:
a (non-self-knotted) Wilson loop labeled $\om$ has expectation $0$ unless no other Wilson lines (except the vacuum, {\bf 1}) link through its center (See Fig.~\ref{KillingHandle}).

\begin{figure}
\begin{center}
\includegraphics[width=0.8\columnwidth]{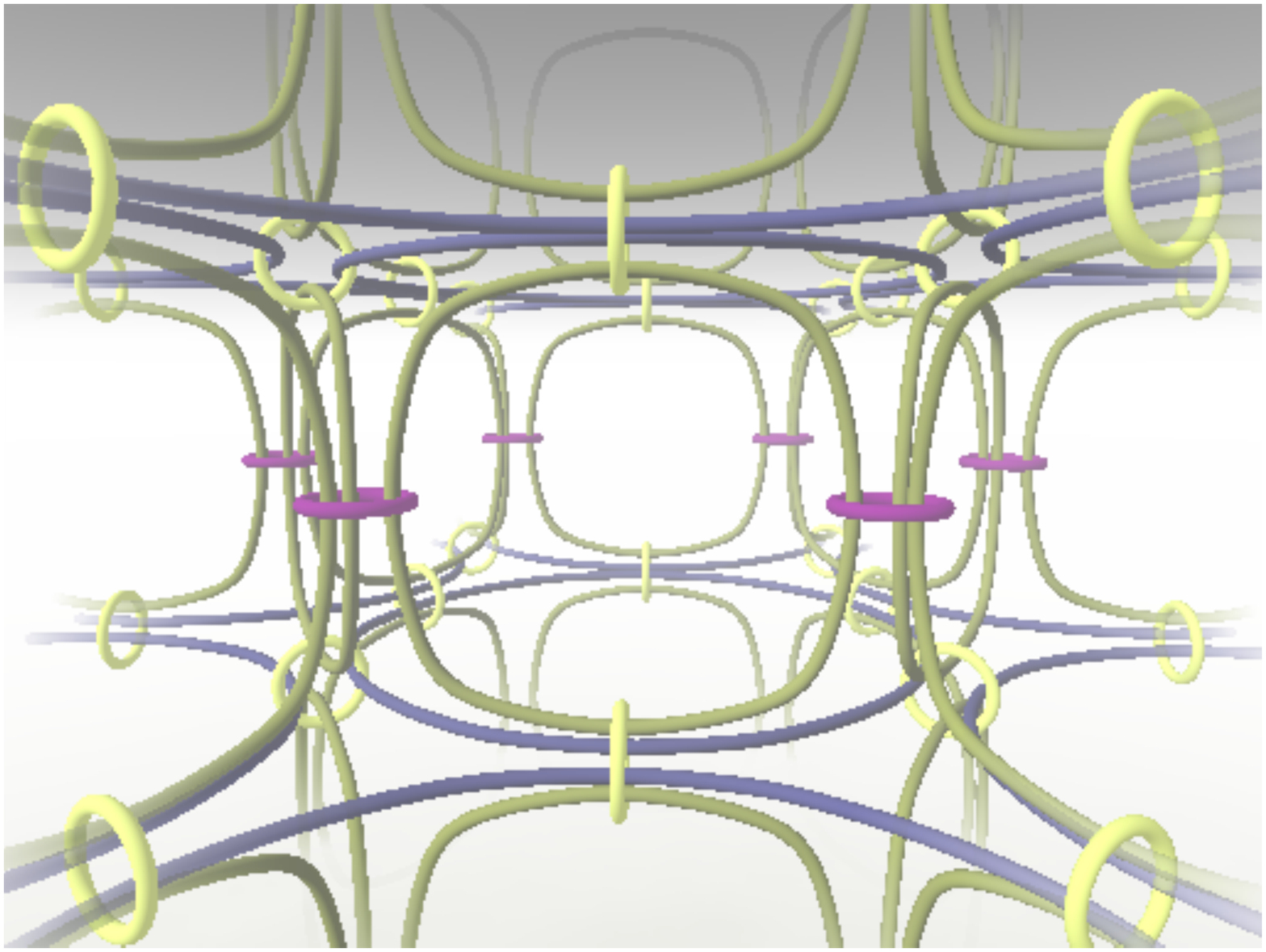}
\end{center}
\caption{The Chain-mail link on a hexagonal lattice.  (Imaginary) time is vertical. Each strand is a Wilson line carrying the sum of quantum numbers $\om$.  Evaluating the expectation of  these Wilson loop operators (the corresponding link invariant) gives the partition function of the LW lattice model.}\label{Chain}
\end{figure}

Armed with this picture of the states and operators in terms of Wilson lines of a CS theory, we now relate the (2+1d) partition function of the LW lattice model to an expectation of linked Wilson loops in a continuum CS theory.
Essentially, we will construct a $3$d link diagram which evaluates the
transfer matrix of the lattice model at successive time steps.
To do this, we first extend our 2d lattice system to a 3d spacetime by reproducing our lattice many times ---  each displaced by a distance $\delta t$ in the (imaginary) time direction -- and adding edges between corresponding points of the lattice in adjacent time slices.
We call a plaquette or edge of this 3d lattice space-like if it is at fixed time and time-like if it extends between two neighboring times.
The label $\bf i$ on each edge at a given time step is represented by a Wilson loop running along this edge, up in time, back along the same edge, then down in time to form a closed loop that traces the  perimeter of the corresponding time-like plaquette (green loops in Fig.~\ref{Chain}).   To apply the vertex projectors at each time step, we encircle the three lines emanating upwards from each vertex by a Wilson line labeled $\om$ (purple loops in Fig.~\ref{Chain}). Since $\om$ is a projector, the labels of the three incident edges must fuse to $\bf 1$, and this projects onto one of the string-net states described above at each time slice.  To apply the plaquette projectors we add an $\om$ loop (blue in Fig.~\ref{Chain}) around the inside perimeter of each space-like plaquette.  In order to force the plaquette projector to act on the string-net states by fusion, we encircle each space-like edge with  another $\om$ string (yellow in Fig.~\ref{Chain}).  The proof that this gives the correct matrix element for the plaquette projector involves some algebra and is given in Ref.~\onlinecite{Us}.  However, it is easy to see that since the plaquette string fuses on each edge with strings at the previous and next time step, it has the desired effect of flipping quantum numbers at each time step $t$.  Finally to obtain the LW partition function we must sum over all possible quantum numbers on the edges (green in Fig.~\ref{Chain}) at all times.  It turns out that this sum can also be effectuated by labeling these (green) Wilson lines with $\om$ which, as mentioned above, is a sum over all possible quantum numbers.  Thus the partition function of the LW model is obtained by evaluating the expectation of the link of Wilson loops shown in Fig.~\ref{Chain} where all loops carry the superposition of labels $\om$.

The diagram in Fig.~\ref{Chain} is in fact known in the mathematical literature as the \ch link; evaluating the expectation values of all Wilson lines gives a topological invariant of the spacetime 3-manifold, called the \ch invariant\cite{Roberts}, which is therefore equal to the partition function of the LW Hamiltonian.  In Appendix \ref{HandleApp}, we give examples of \ch links for $S^2 \times S^1$ and $T^2 \times S^1$, which can be represented in a manner which locally resembles Fig. ~\ref{Chain}.  If we are concerned with lattice models on a 2d manifold M, then our resulting 3-manifold will necessarily be M x S1 (with compactified time).  However, the chainmail construction is not limited to product manifolds of this type.

A priori, it may seem that we have turned the comparatively simple problem of evaluating the partition function of a lattice Hamiltonian into the evaluation of a very complicated set of inter-linked Wilson lines.  However, $\om$ has some rather remarkable properties which render this evaluation considerably simpler than it appears.
The first is known as the ``handleslide" property, and essentially follows from the fact that $\om$ is a projector (See Fig.~1): a line bearing {\em any} label can freely ``slide over" an $\om$ loop without changing the expectation value of the diagram\cite{Roberts,handleslide,Wang}.  By handlesliding the various Wilson lines in Fig.~\ref{Chain} over each other,
and using the $\om$ to fuse strings together (See Fig.~1), it is possible to eliminate all but a (typically small) finite number of
strings
from the diagram
\cite{Roberts}.  Hence a combinatorially complex diagram can be reduced to one in which only a small number of matrix elements must be computed.
This is precisely the partition-function analogue of the tensor network renormalization procedure known to be exact for LW Hamiltonians\cite{LevinNaveWenGu}.
 Similarly one may
prove that the value of the \ch invariant is independent of the lattice structure\cite{Roberts}.
This lattice independence immediately shows how we may construct LW models on arbitrary 2d lattices.
To do so, we follow the same prescription as in the honeycomb case, extending the 2d lattice to 3d, adding $\om$ loops around all the perimeter of all plaquettes, and wrapping an $\om$ around each lattice edge encircling the corresponding plaquette strands. The resulting model has identical topological properties as that built on the honeycomb lattice, but the Hamiltonian has a slightly different form due to the different geometry (See Ref.~\onlinecite{Us}).

The second remarkable property of the $\om$ loop is that the expectation value of any collection of linked $\om$ loops is known to be\cite{Witten,Wang}  equal to the {\em vacuum} partition function $Z_{CS}$ of a CS theory evaluated in a spacetime manifold which is topologically different from the one in which the $\om$ loops reside.  Hence we may relate the
partition function of our lattice model, which can be evaluated as an expectation of $\om$-labeled Wilson lines, to the vacuum partition function of a continuum CS theory in a different spacetime.
The precise procedure for altering the topology of the spacetime manifold to remove the $\om$ loops is known as ``surgery".  Applied to the \ch link\cite{Roberts,endnote3}, surgery maps the spacetime $\cal M$ on which the lattice lives  to a new manifold ${\cal M} \# \overline {\cal M}$, meaning the manifold $\cal M$ connected to its mirror image $\overline {\cal M}$.  This explains a crucial fact about the lattice models: although all of the Wilson lines are from the same chiral CS theory, the LW ground state partition function $Z_{LW}$ is achiral:  $Z_{LW} = Z_{CS}({\cal M} \# \overline {\cal M}) = |Z_{CS}({\cal M})|^2$.   To summarize, we map the lattice model to $\om$ loops in the spacetime manifold ${\cal M}$, then we use surgery to eliminate all $\om$ loops, resulting in a theory that lives on a spacetime manifold consisting of {\em two} copies of the original spacetime $\cal M$-- one with the same, and one with the opposite, chirality.

We now consider adding quasi-particle (qp) excitations  to the partition function described above.    In the Hamiltonian picture, these qps ``violate" either the vertex or the plaquette projectors, or both, incurring a corresponding energy penalty $E_G$.  (The purely topological CS partition function cannot capture the $e^{-E_G \tau}$ contributions; instead qps in this language correspond to external sources of the gauge field.)  To violate the vertex projector $V_m$, we must force the Wilson lines entering the vertex $m$ to fuse to some non-trivial representation $\bf i\neq 1$.  This is done by threading an additional string labeled with the conjugate representation (or antiparticle) $\bf i^*$ through the vertex projector loop.
Similarly, we can violate a plaquette operator by passing a labeled string through the corresponding plaquette projector loop.
Thus a qp world-line corresponds in the spacetime diagram to a Wilson line carrying the representation $\bf i^* \neq 1$ which passes through the plaquette and/or vertex loops.

In LW models based on a CS theory with $\bf r$ different particles, there are ${\bf r}^2$ species of qps, which are the tensor product of two noninteracting sectors of $\bf r$ qps with opposite chiralities\cite{LevinWen}.
Hence we can understand the full spectrum by understanding how to add qps from each of these two sectors.  Qps in the first, ``right handed" (RH) sector, are Wilson lines that follow the edges of the lattice, passing through the $\om$ loop encircling each edge.  In the ``left handed" (LH) sector, the qp world lines again follow the edges and go through the corresponding $\om$ loops, but when the LH qps pass between $3$-cells of the 3d lattice, they must detour through an $\om$ wrapping around the perimeter of a plaquette.  (See also Ref.\cite{Trebst} for an alternate description of the same conclusion).  These two species of qps are shown in Fig.~3.  Since for both the RH and the LH qp, each world line can be labeled with $\bf r$ different quantum numbers, we obtain the full set of ${\bf r}^2$ species for the doubled theory.

\begin{figure}
\begin{center}
\includegraphics[width=0.7\columnwidth]{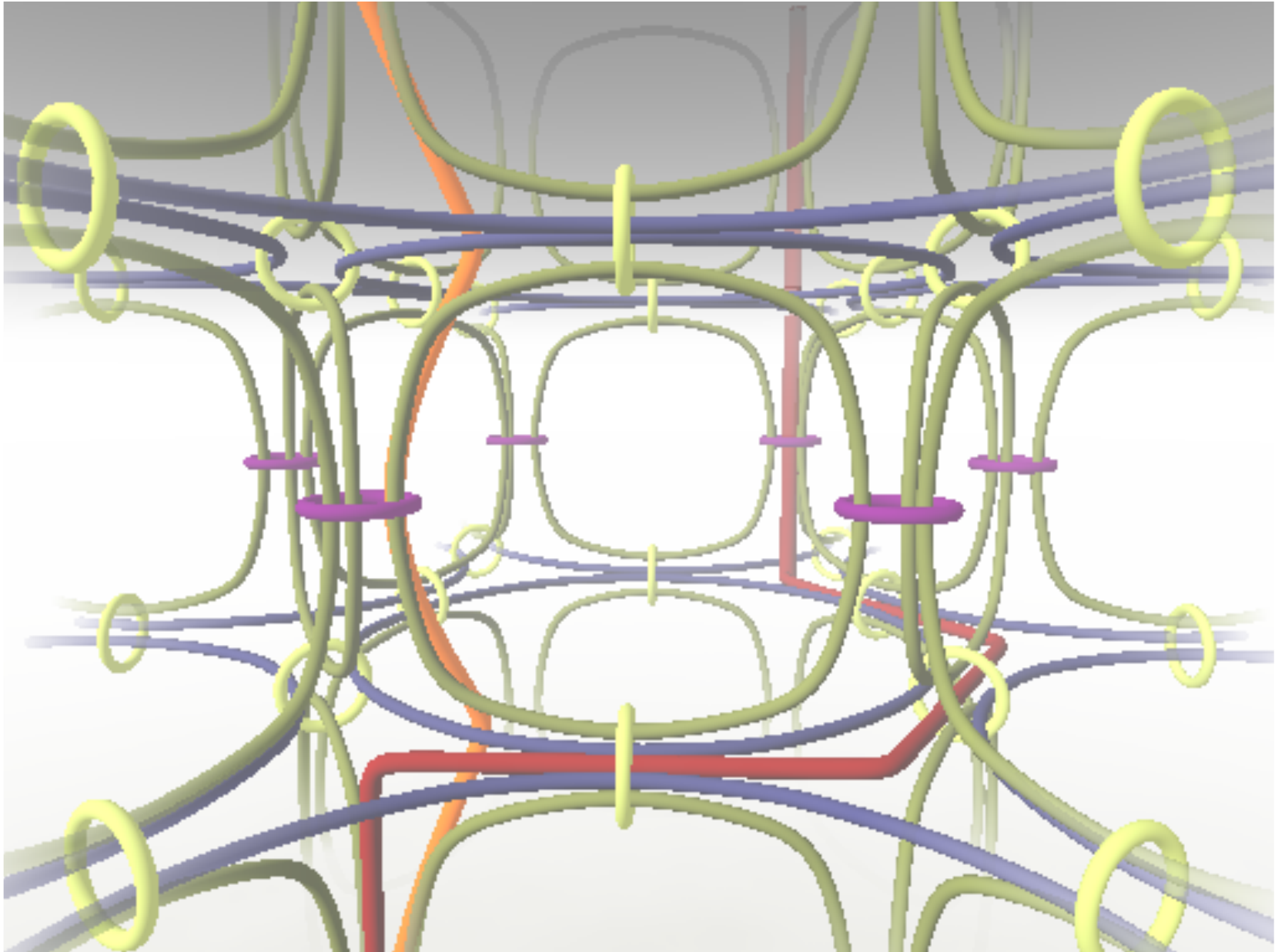}
\includegraphics[width=0.7\columnwidth]{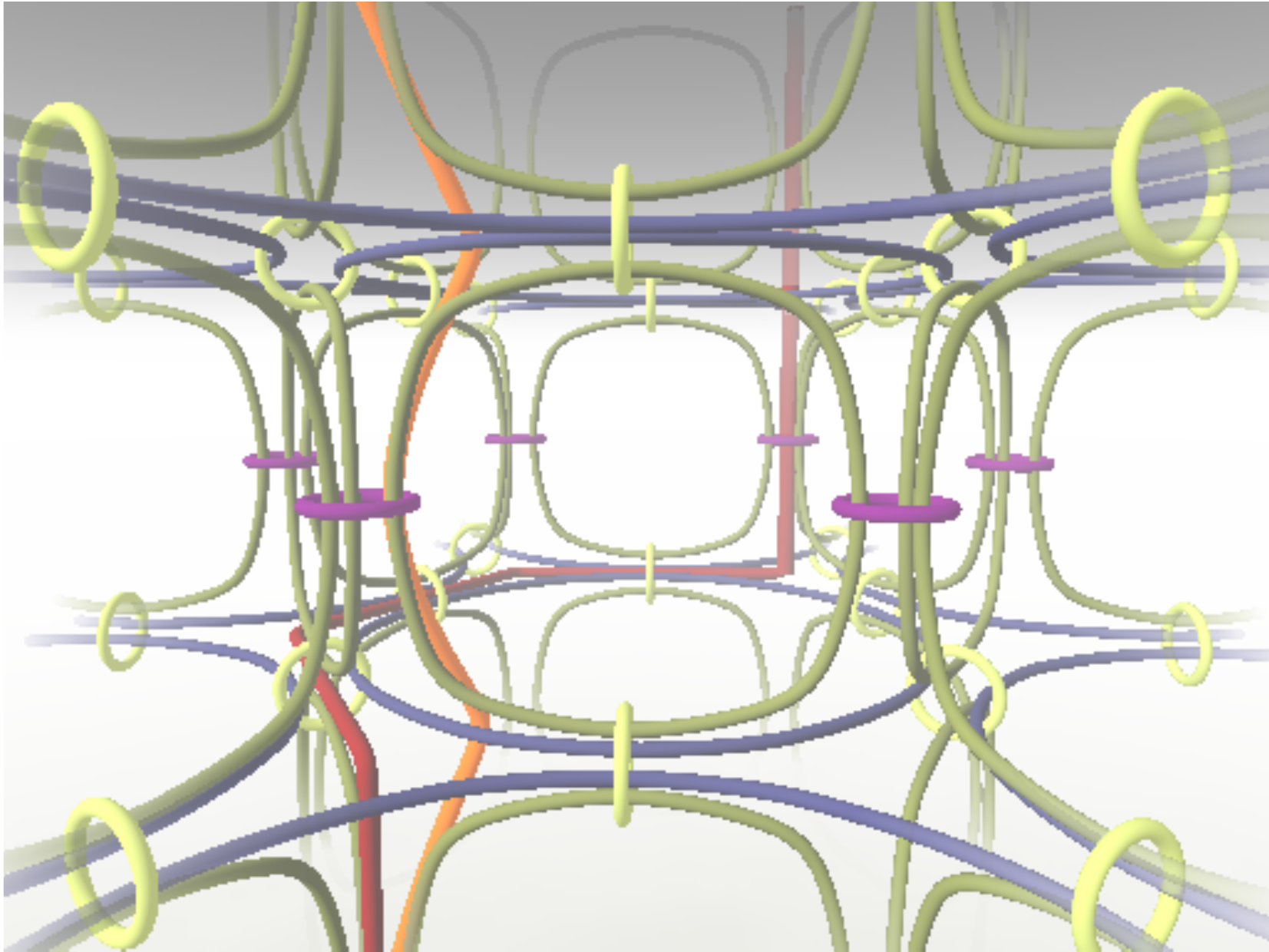}
\end{center}
\caption{(Top) LH (orange) and RH (red) quasi-particles on the \ch link.  LH qps link both edge (purple and yellow) and plaquette (blue and green) $\om$ loops; RH qp's link only the edge loops.  (Bottom) The RH qp after handle-sliding over a (horizontal blue) plaquette loop.  Such slides allow RH and LH qps to pass freely through each other, demonstrating that they have trivial mutual statistics.}\label{QPS}
\end{figure}

{\it We can now deduce the statistics of the quasi-particle world-lines, as well as their contribution to the partition function in the case that they do not enclose non-contractible loops in $\mathcal{M}$.}
Using the handleslide property of the $\om$ loops, the qp world lines can be geometrically deformed (See Fig.~3).
As shown in the figure an RH and an LH world line can freely handleslide through each other, showing that these are separate non-interacting sectors.  However, an RH cannot pass through an RH nor can a LH pass through an LH, so that as expected these qps have nontrivial braiding statistics within their own sectors.  The detailed statistics of each sector can also be understood using handleslides to separate these qps from the \ch link entirely\cite{Us}; the RH 
sector has the same statistics as the CS theory used to construct the lattice model, while Wilson lines in the LH sector {\it reverse} their chirality as we un-link it from the $\om$ loops in the \ch link  (Details in Ref.~\onlinecite{Us}). Hence 
although all qps are inserted by means of Wilson lines of a single chiral CS theory, the lattice model nonetheless contains two species of qps with opposite chirality.{\it Sliding the qp world-lines off the link in this way also shows that 
$Z = Z_{VAC} \langle \mathcal{L} _{QP} \rangle $, where $Z_{VAC}$ is the partition function in the absence of quasi-particles, and $\langle \mathcal{L} _{QP} \rangle $ is the Kauffman bracket of the quasi-particle world-line link.  }

The appearance of this reversed chirality can also be understood within the surgery description of the Chain-Mai link.  As discussed above, after surgery on $\om$'s, the spacetime manifold is ${\cal M} \# \overline {\cal M}$.
As shown in Ref.~\onlinecite{Martins}, after surgery, RH world lines end up within the ${\cal M}$ manifold and thus retain the structure of the original CS theory.  However, extending the work of Ref.~\onlinecite{Martins} we find that the LH qps instead end up in the $\overline {\cal M}$ manifold and thus obtain similar structure but with reversed chirality.

This work has shown how certain LW lattice models are related to a continuum CS description, giving insight into how the theory becomes doubled.  Further, we give a simple visualization of topological invariance, making
coarse-graining and independence of lattice structure in these models manifest.  Although this paper has focused on CS theories, it can be almost trivially generalized\cite{Us} to describe LW models constructed from any system of anyons (or ``modular tensor category").

\begin{appendix}

\section{Appendix: Examples of Chain-Mail links for simple $3$-manifolds}  \label{HandleApp}

In this Appendix we give  examples of  \ch links appropriate  to some simple closed spacetime $3$-manifolds.  
Given a space-time $\mac{M}$, we first describe a prescription to construct a simple  \ch link $L$.  The partition function (at infinite quasi-particle mass) is given by
\be \label{Eq_Z}
\mac{Z} = \mac{D}^{-n_0 -n_3} \langle L_{\Omega} \rangle \ \ \ .
\ee
where $L$ is a \ch link of Wilson lines.  Here the notation $L_\Omega$ indicates that all Wilson lines are labeled by the superposition of representations $\Omega$.  To evaluate the partition function, we use the result\cite{WittenJones} that $\langle L_{\Omega} \rangle$ is given by a link invariant (closely related to the Jones polynomial when the gauge group is $SU(2)$) of the link $L$ of Wilson lines.  (The pre-factor $\mac{D}^{-n_0 -n_3}$ is a normalization; we will define $n_0$ and $n_3$ shortly).  There are many possible choices of \ch link for each $\mac{M}$ -- a fact which we can exploit to `decorate' the simple \ch link and arrive at a link describing a lattice model.

\subsection{Handle Decompositions}

\begin{figure}[h!]
\begin{center}
\includegraphics[width=0.95\columnwidth]{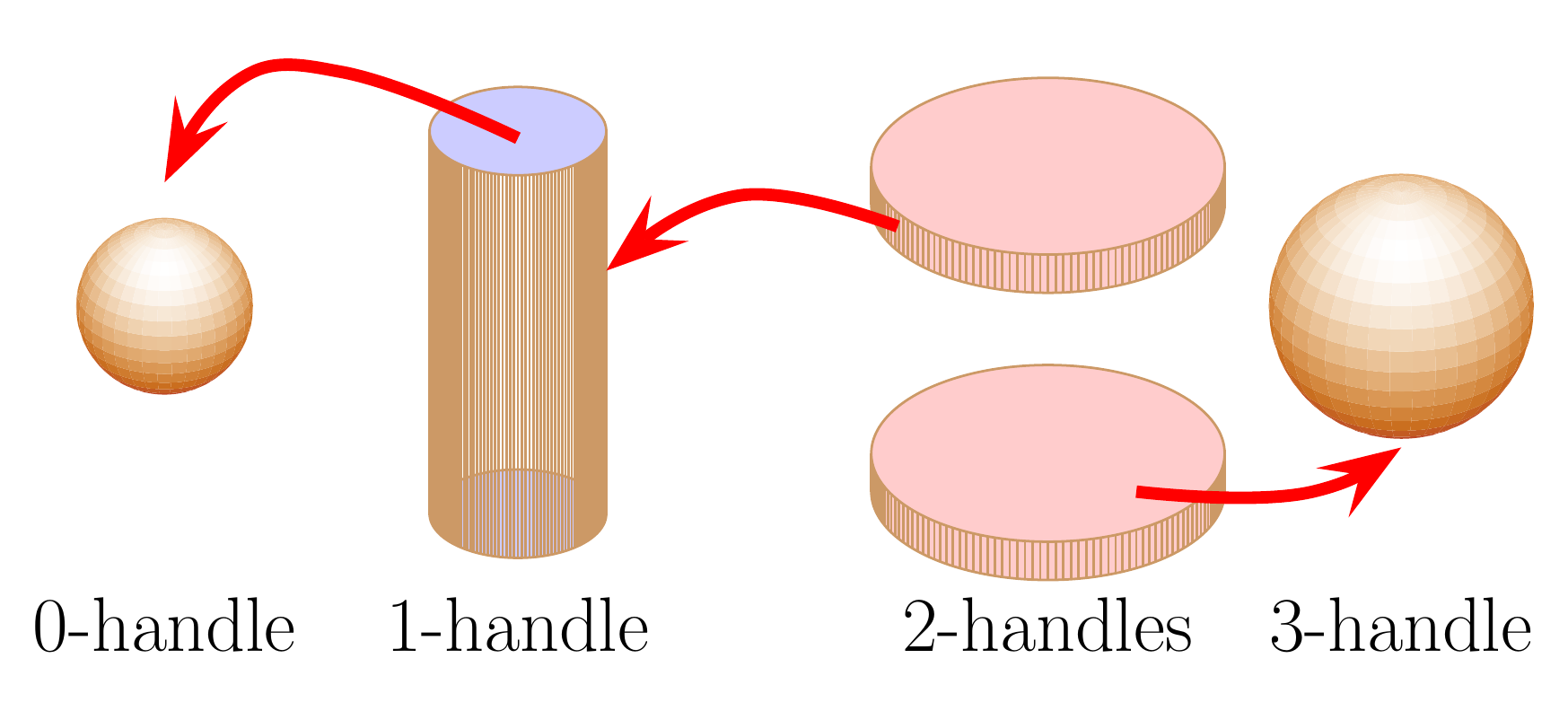}
\caption{ \label{Fig_Handles} The four types of handles used to construct general $3$-manifolds.  The $0$-handle will be the template onto which other handles are glued.  The $1$-handle is a tall cylinder, which we attach along its ends (shown here in mauve).  The $2$-handle is a short cylinder, which is glued in along its circumference (shown in brown).  The $3$- handle is a solid ball, which is glued in along the sphere which bounds it.}
\end{center}
\end{figure}

To construct a \ch link, we use a {\it handle decomposition} of $\mac{M}$\cite{NoteA}.   The general idea is to construct $\mac{M}$ by gluing together a collection of {\it handles}, each of which is, in isolation, topologically equivalent to a solid ball.  There are four different types of handles that we will need, distinguished by the rules that we will use to glue them in (Fig. \ref{Fig_Handles}).
We begin with an array of $0$-handles (or points).  To these we may attach $1$-handles (best visualized as long, thin solid cylinders) by glueing them onto $0$-handles along the two disks at the end of the cylinder.
We next attach $2$-handles (short, fat solid cylinders) by glueing them onto $0$ or $1$ - handles along the circumference of the cylinder.  Finally we fill in the empty space by gluing 
 $3$-handles (solid balls) onto $2$-handles along the sphere bounding the ball.  
%Topologically speaking, $1$-handles are effectively glued onto points (which are in fact the $0$-handles), $2$-handles are glued onto circles (which are obtained by gluing $1$-handles onto $0$-handles), and $3$-handles are glued onto spheres (which are obtained by gluing in $2$-handles).  Thus to construct our manifold, we start with some $0$-handles, then glue in $1$-handles, and so on in order of increasing handle index.  (As we shall see it is possible to skip an index entirely, however, in which case for example a $2$-handle may be attached directly to a $0$-handle).
In Eq. (\ref{Eq_Z}), $n_0$ and $n_3$ are the number of $0$- and $3$- handles, respectively.

To illustrate, we consider two examples: $S^2\times S^1$ and $T^2\times S^1$.  Here we view $T^2$ and $S^2$ as describing space, and $S^1$ as describing compactified imaginary time.  (We can view this as computing the partition function at finite temperature.  The result gives the lattice partition function at infinite quasi-particle mass).    

\subsubsection{$S^2 \times S^1$}

\begin{figure}[h!]
\begin{center}
\includegraphics[width=0.8\columnwidth]{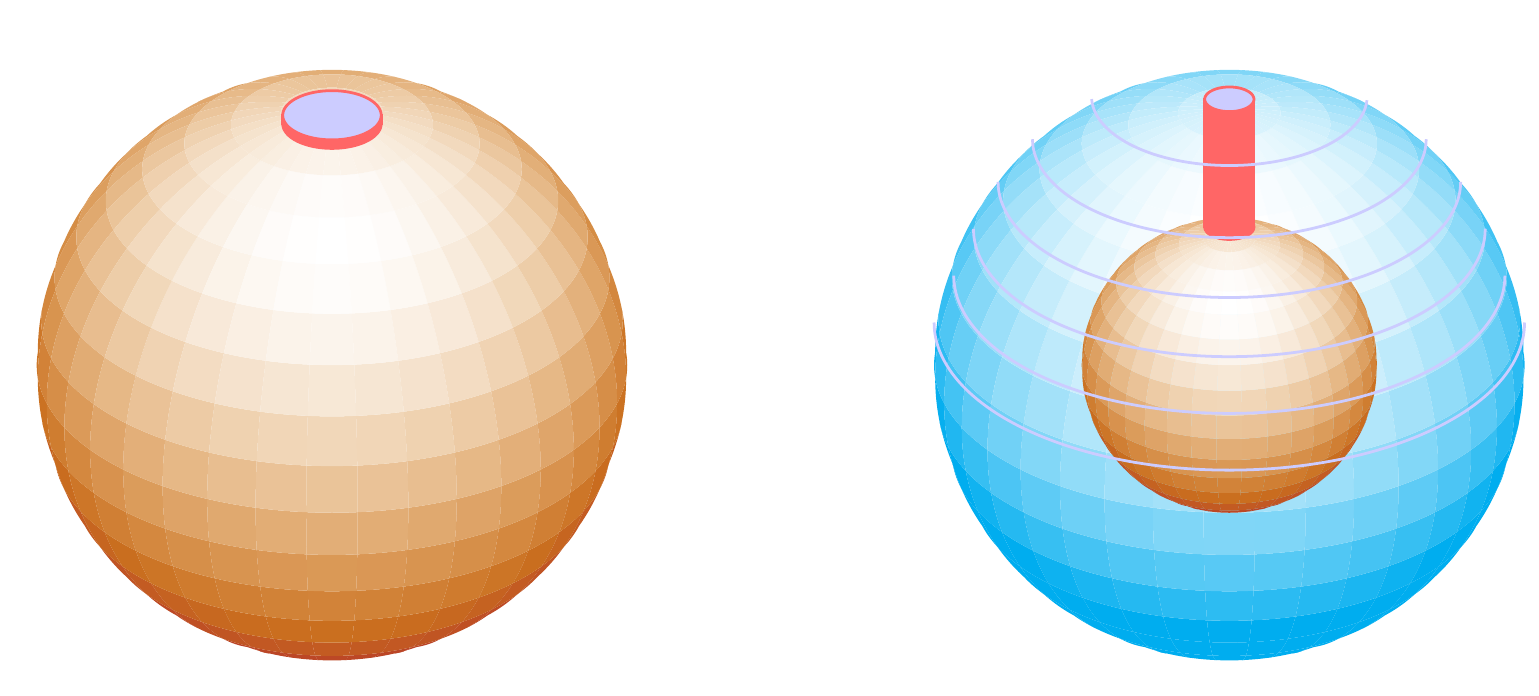}
\caption{ \label{Fig_S2I} A simple handle decomposition of $S^2 \times S^1$.   (a) A simple handle decomposition of $S^2 \times I$, consisting of a single $2$-handle (brown) attached to a single $0$-handle (mauve) along the boundary of a disk (pink).  (b) To make $S^2 \times S^1$, we take two concentric copies of (a), joined by a $1$-handle in the (radial) time direction.  To compactify time, the two $S^2$ space-like surfaces are identified.   }
\end{center}
\end{figure}

First, we give a handle decomposition of $S^2 \times I$.  Many readers will be familiar with the fact that one can map a plane onto a sphere by stereographic projection, which maps the origin onto the sphere's North pole, and the circle of points at infinity to the sphere's South pole.  This is essentially the idea that we will use to construct a sphere: we will begin with a $0$-handle and a $2$-handle.  To make a sphere, we glue the circumference of the $2$-handle cylinder onto the $0$-handle, as shown in Fig. \ref{Fig_S2I}.  The result is $S^2 \times I$, since the $0$- and $2$-handles are also thickened in the third dimension.

The next step is to identify the end-points of the interval $I$ to turn $S^2 \times I$ into $S^2 \times S^1$.  
 To do this, we take two copies of the existing handle decomposition ($\mac{H}_0$), and join them by attaching a $1$-handle that runs between the two copies of the $0$-handle.  This gives a cylinder joining two concentric spheres (Fig. \ref{Fig_S2I}b).  The boundary of the hollow space between the two spheres is topologically a sphere, which we can see by shrinking the cylinder and inner sphere to a point.  Hence we may fill it in by gluing in a single $3$-handle, which is attached along this spherical boundary.  
To obtain a closed $3$-dimensional space-time,  we identify the two remaining spherical boundaries (one from each copy of $\mac{H}_0$).  The net result is a handle decomposition with one $0$-handle $h_0$, one one-handle whose end-points are glued to $h_0$, one two-handle, which is also attached to $h_0$, and one $3$-handle.

\subsubsection{$T^2 \times S^1$ }

 Our second example is a spatial torus  with compactified imaginary time.  We  begin by building $T^2 \times I$, which is a finite thickened region of the plane with periodic boundary conditions.   This requires one $0$-handle (representing the $4$ vertices of the square, all of which are identified), two $1$-handles (one for the horizontal, and one for the vertical, edges), and a $2$-handle for the face.   
 The $2$-handle attaches to the $1$-handles by first tracing forward along $a$, then forward along $b$, then backward along $a$, then backward along $b$.  A convenient short-hand for this is to say that the $2$-handle attaches along the trajectory $aba^{-1} b^{-1}$.  

\begin{figure}[h!]
\begin{center}
\begin{tabular}{c}
\includegraphics[width=0.9\columnwidth]{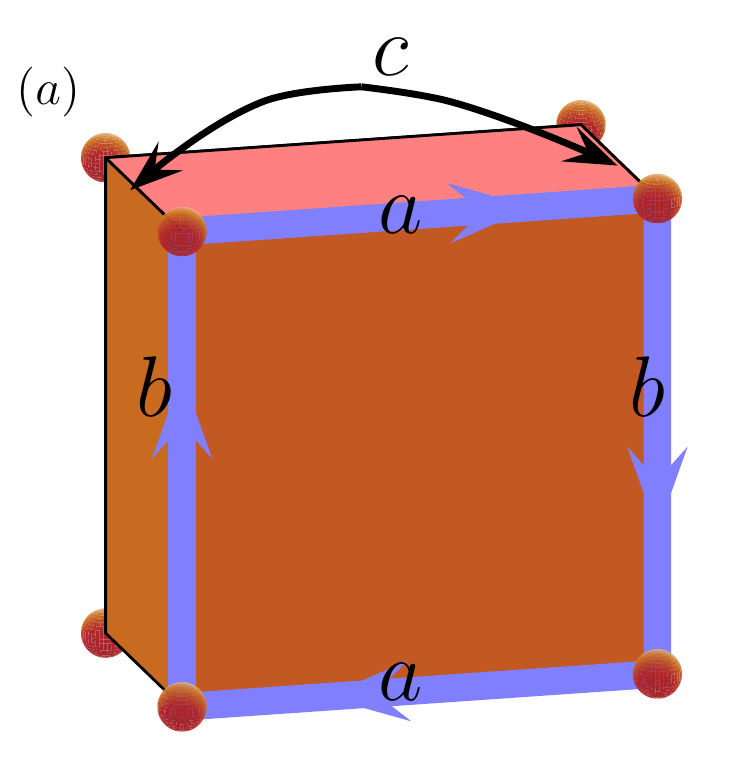} \\
\includegraphics[width=0.9\columnwidth]{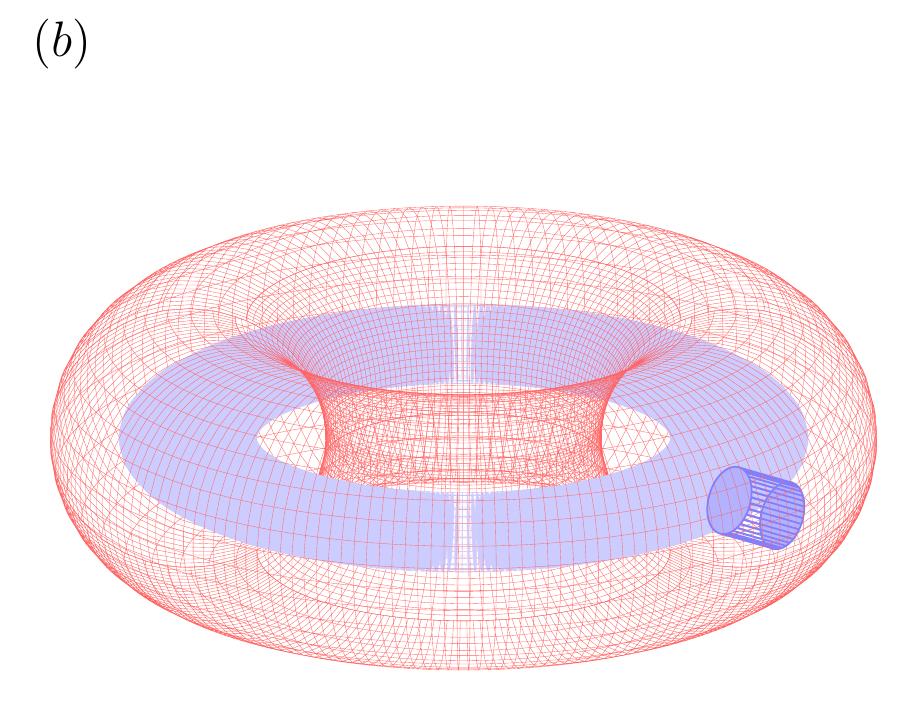}
\end{tabular}
\caption{ \label{Fig_T3}  A handle decomposition of $T^2 \times S^1$.  (a) shows the `unfolded' handle decomposition, with arrows indicating the trajectory of the $2$-handle attaching maps.   To obtain $T^2\times S^1$, we must identify all eight corners of the cube, each quadruple of edges labeled by the same letter ($a$, $b$, and $c$), as well as the pairs of faces which are opposite each other on the cube.  (b) shows the decomposition with all of the identifications except those between the top and bottom of the cube.  The result is concentric tori, joined by a single $1$-handle and a pair of $2$-handles (one for each non-contractible curve on the torus; these are not shown in the Figure).}
\end{center}
\end{figure}

To construct $T^2 \times S^1$ from $T^2 \times I$,
we again take two copies of handle decomposition of $T^2 \times I$, and add a $1$-handle in the time direction.  This gives two concentric thickened tori joined by a cylinder (Fig. \ref{Fig_T3}b).  Now, however, we run into a complication: the hollow space between the two tori is not a sphere, it is a solid torus!  Hence before gluing in the $3$-handle, we must add a pair of $2$-handles which will fill in the non-contractible curves inside the torus, such that we can close the space-time by gluing in a $3$-handle.  

The simplest way to visualize this is to draw the handle decomposition on a cube (Fig. \ref{Fig_T3}(a)). Here $a$ and $b$ correspond to the $1$ handles we needed to build $T^2$, and $c$ is the new $1$-handle in the time direction.   In this diagram, $c$ appears along $4$ edges of the cube, as all $4$ of these actually depict the same edge.  Now it is easy to see that we must add two $2$-handles -- one along the top face of the cube (which is identified with the bottom face), and one along its right face (identified with the left face).  Adding these (and accounting for the identifications) fills in all of the empty plaquettes which were created when we added $c$.  The remaining hollow space is simply the center of the cube -- whose boundary is again topologically a sphere, so that we may fill it in with a single $3$-handle.

The net result is a handle decomposition with one $0$-handle $h_0$, three one-handles, all of which are glued at both endpoints to $h_0$, three $2$-handles, and one $3$-handle.  The $2$-handles are attached along the faces of the cube defined by the $1$-handles, tracing the curves $aba^{-1}b^{-1}$, $aca^{-1}c^{-1}$, and $bcb^{-1}c^{-1}$, respectively.

\subsection{Drawing the \ch link}

To compute the \ch invariant (and hence $\mac{Z}_{LW}$ on the spacetime $\mac{M}$), we associate a \ch link to these handle decompositions and evaluate the invariant (\ref{Eq_Z}).   $\mac{Z}$ is the same for any handle decomposition of $\mac{M}$\cite{RobertsThesis}.  
In both examples above, we have chosen decompositions with a small number of handles.  This is convenient for computing the partition function.  Handle decompositions are far from unique, however -- any spacetime $\mac{M}$ can be constructed in many different ways.  We will see that this allows us to relate $\langle L_{\Omega} \rangle$ for the simple decompositions used here to $Z_{LW}$, which as discussed above is related to a complicated link diagram locally resembling Fig. \ref{Chain}.

The prescription for constructing $L_{\Omega}$ is as follows\cite{RobertsThesis}.  First, draw a string which runs around the edge of each $2$-handle, tracing the line along which it was glued onto the $0$ or $1$-handles.  Second, draw a string which runs around the circumference of each $1$-handle, being sure to encircle any $2$-handle strings which run along the $1$-handle in question.  

For example, our decomposition of $S^2 \times S^1$ contains one $2$-handle, which is attached to the $0$-handle. So we begin by drawing a loop along the $0$-handle (a closed loop around one end of the cylinder in Fig. (\ref{Fig_S2I}).  The decomposition also contains one $1$-handle, though the $2$-handle is not attached along it.  Hence we draw a second loop, encircling the circumference of this $1$-handle (around the middle of the cylinder in Fig. \ref{Fig_S2I}), which is not linked with the first.  Evaluating the link invariant (using the rules outlined in Appendix \ref{knotApp}) gives 
\be
\mac{Z}_{S^2 \times S^1} = \frac{1}{\mac{D}^2} \langle \includegraphics[height=.1in]{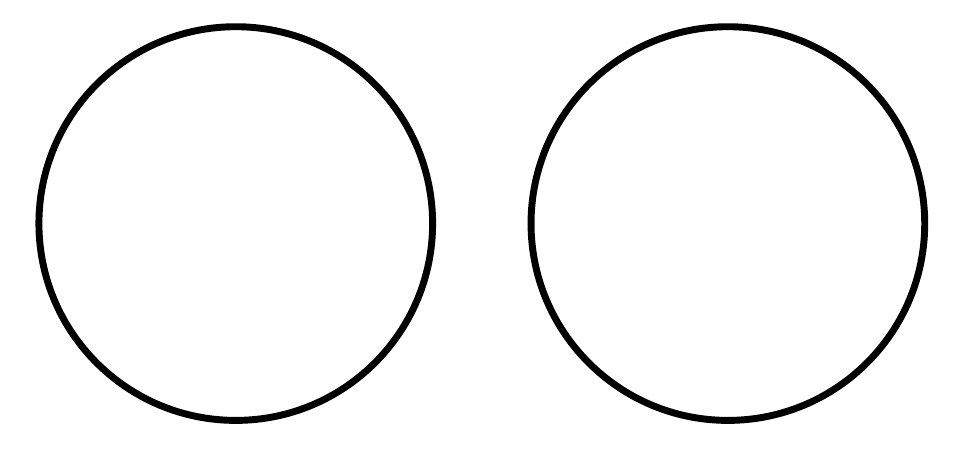} \rangle = 1
\ee

\begin{figure}[h!]
\begin{center}
\includegraphics[width=0.8\columnwidth]{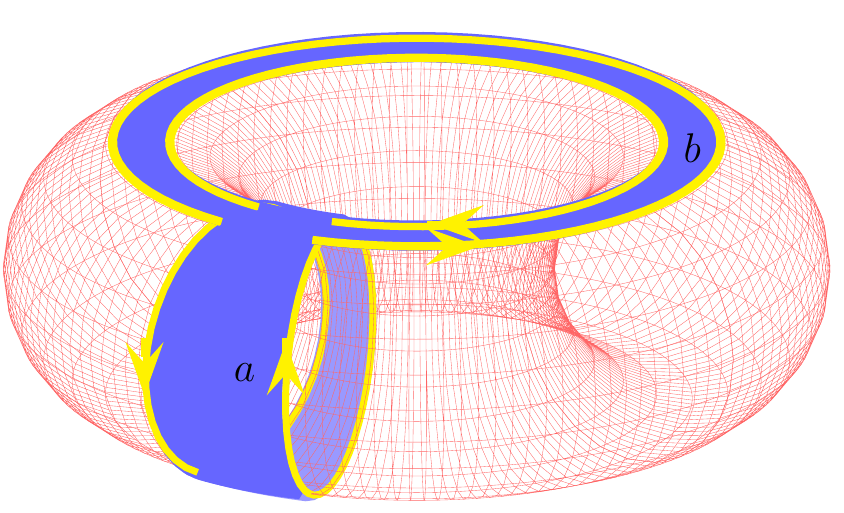}
\caption{ \label{Fig_T2CH}  An illustration of how to draw the $2$-handle string to correctly construct the torus.  The $1$-skeleton of $T^2$ consists of two strips ($a$ and $b$) joined at a point.  The $2$-handle string (shown here in yellow) begins at the crossing of the two one-handles, runs forwards around the right edge of $a$, forwards around the bottom edge of $b$, backwards around the left edge of $a$, and finally backwards around the top edge of $b$.  }
\end{center}
\end{figure}

The \ch link for $T^2 \times S^1$ is slightly more involved.  
We first consider the torus, as the result is easier to visualize (Fig. \ref{Fig_T2CH}).  After identifying the appropriate pairs of $1$-handles on the square, the $1$-skeleton of the torus is simply two rings joined at a point.  We may flatten these rings a bit, so they look like two circular strips, labeled $a$ and $b$.  To attach the $2$-handle onto this picture, we start at the $0$-handle where these strips cross, and proceed along the right edge of $a$.  Once the string has gone all the way around $a$, we trace it along the bottom edge of $b$.  When this circle is complete, we trace it along the left edge of $a$, being sure to go in the opposite direction as last time.  Finally we trace along the top edge of $b$, again in the opposite direction as before.  This attaches the boundary of the $2$-handle along $aba^{-1}b^{-1}$.  

\begin{widetext}

\begin{figure}[htp]
\begin{center}
\begin{tabular}{cc}
\includegraphics[width=0.4\columnwidth]{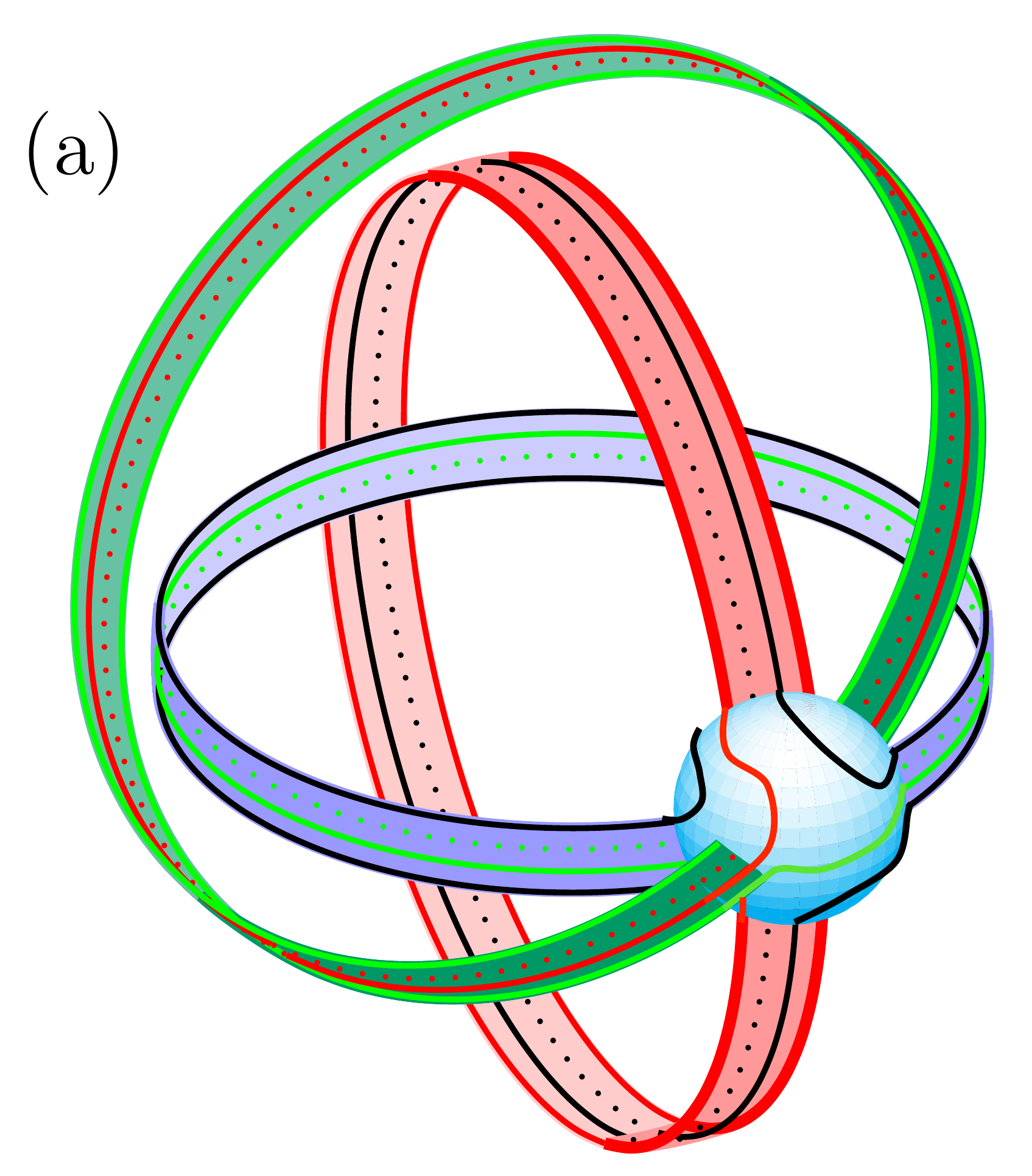}& \ \ \ \ \ \ \ \ \ \
\includegraphics[width=0.4\columnwidth]{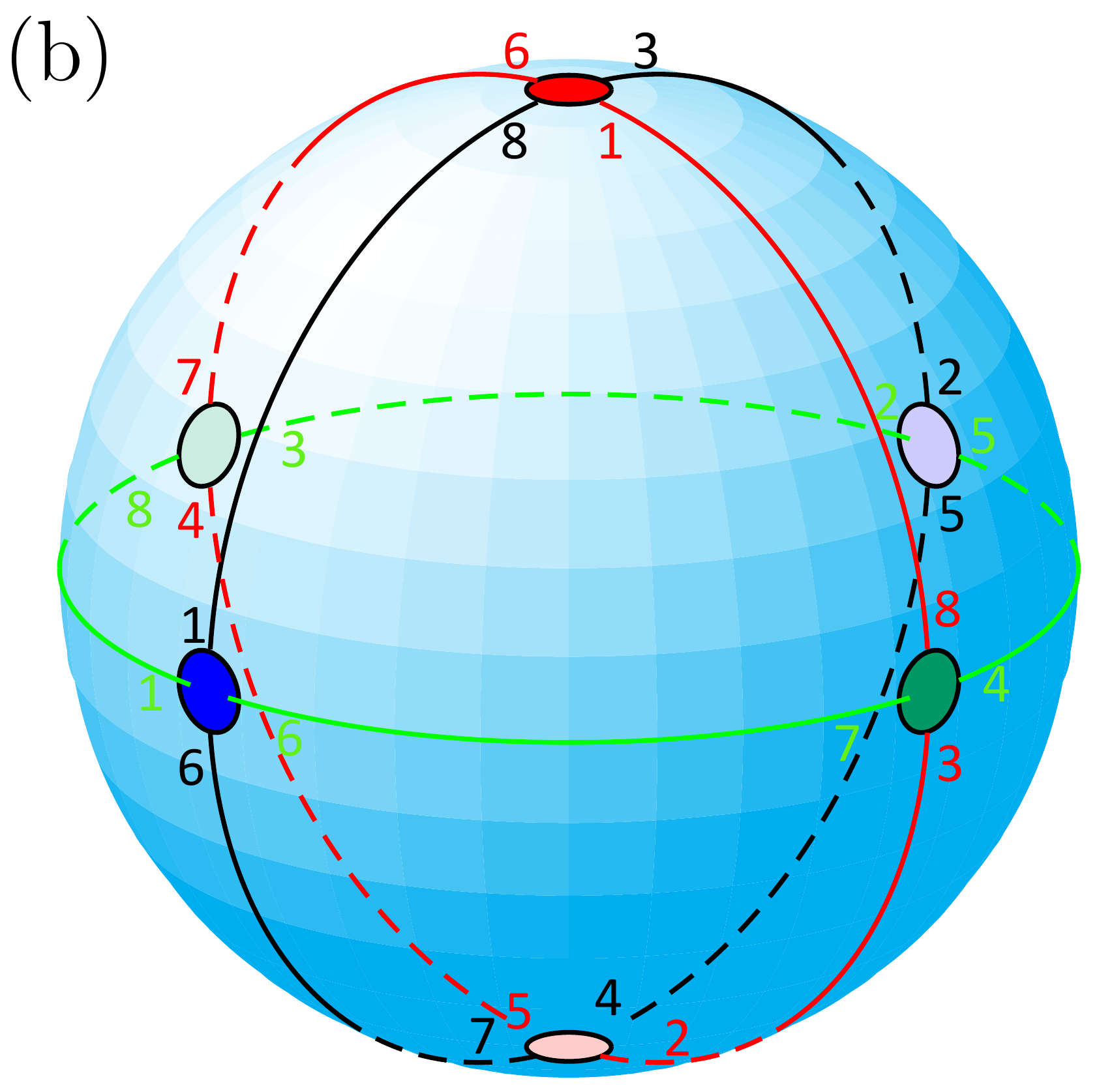} \\
\includegraphics[width=0.4\columnwidth]{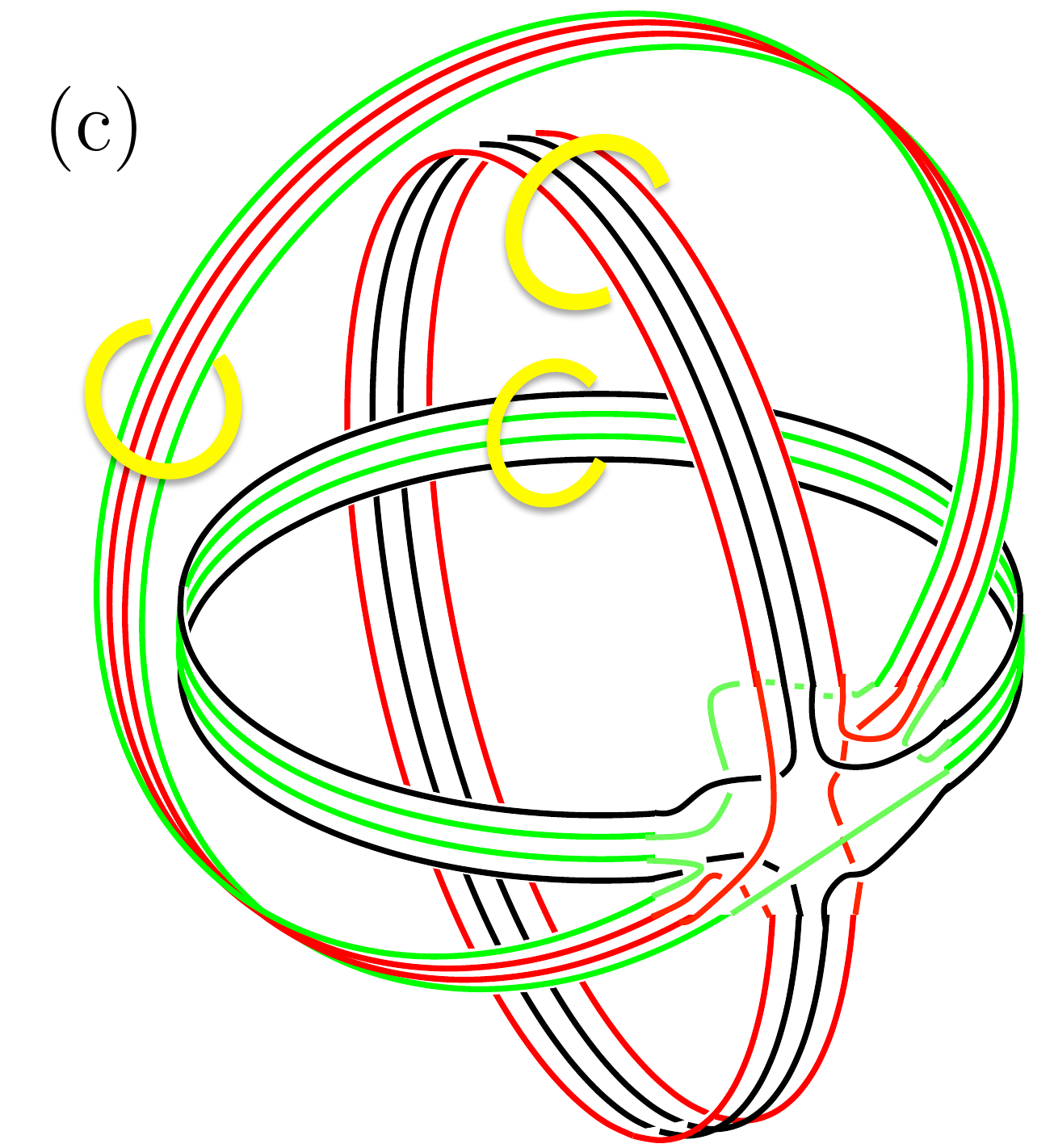}& \ \ \ \ \ \ \ \ \ \
\includegraphics[width=0.4\columnwidth]{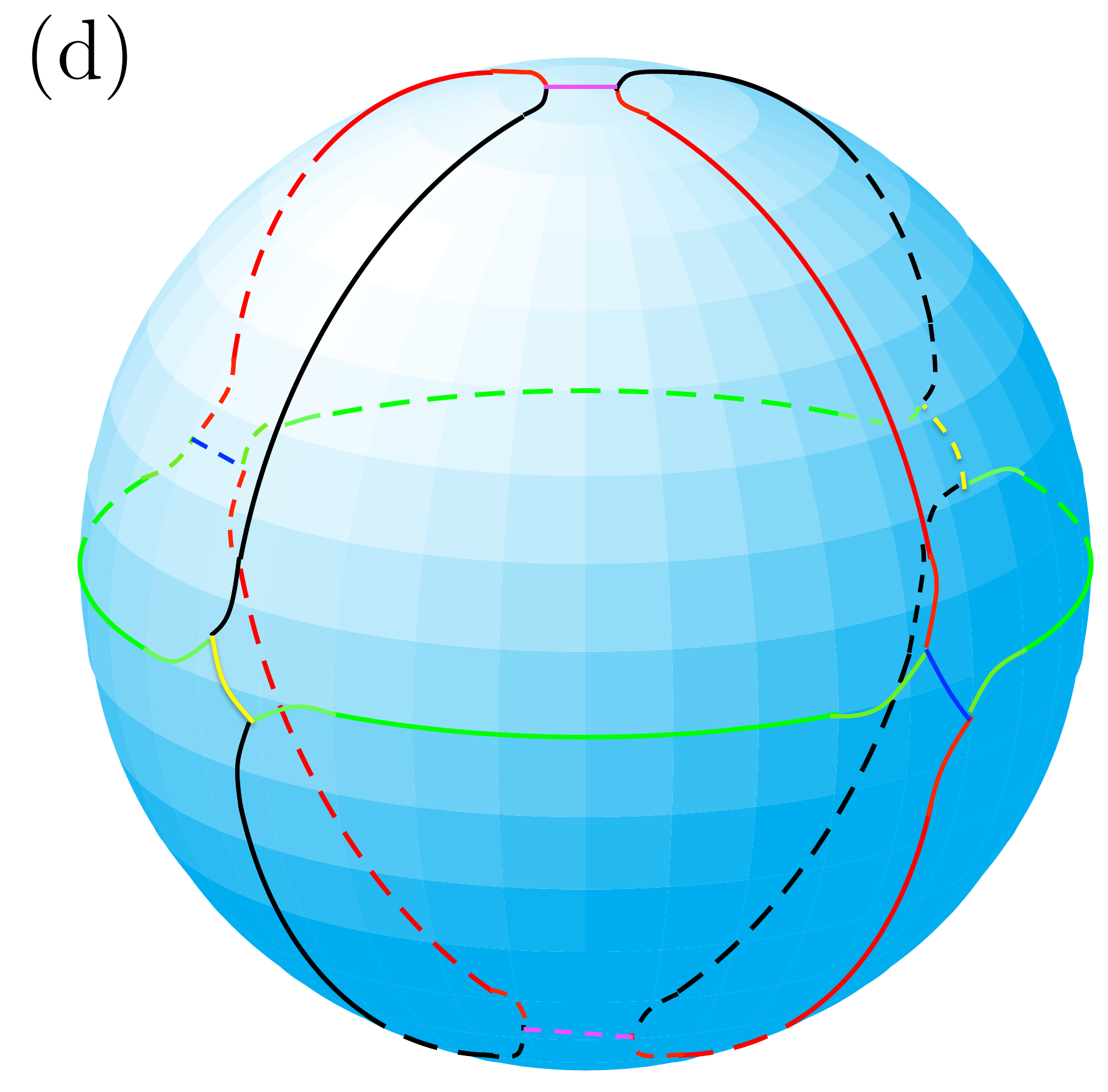} \\
\end{tabular}
\caption{ \label{Fig_T2S1CH}  The attaching maps of the $2$-handles to the $1$-skeleton of $T^2 \times S^1$. (a)  The $1$-skeleton of $T^2 \times S^1$ consists of one $0$-handle $h_0$, represented here as a ball, and three $1$-handles ($h_a$ in green, $h_b$ in blue, and $h_c$ in red) which are attached to its spherical surface.  We may think of the $1$-handle attached at the top and bottom of the sphere ($h_c$) as the time-like direction, and the handles attaching to the front and back, and left and right, ($h_a$ and $h_b$) as the two non-contractible curves on $T^2$.  The attaching maps of the 2 handles are drawn with solid lines when they are on the side of the $1$-handle facing out of the page, and dotted lines otherwise.  (b) The attaching maps of the $2$-handles without the $1$-handles.  Each attaching map is drawn in a different color; the numbers represent a sequence of points through which the map of this color must pass in order.  For example, the $2$-handle which comprises the space-like $T^2$ begins at the point (labeled by a green number $1$ in the figure) where $h_a$ is attached to $h_0$.  It traces straight around the top of $h_a$ to emerge at the antipodal point on the sphere (green number $2$).  It then crosses the surface of the sphere to the left-most end- point of $h_{b}$, and follows $h_b$ around to the antipodal point ($3$ and $4$, respectively).  From here it returns to $h_a$, starting at the back of the sphere this time ($5$) and returning to the front ($6$).  Finally it makes its way to the left-most end-point of $h_b$ ($7$) and follows $h_b$ around to the leftt side of the sphere ($8$).  From here, it returns to its starting point.   The trajectories of the other two strings are similar, as shown.  (c) shows the resulting chain-mail link.  To evaluate this, we enforce the condition that no flux passes through the yellow rings.  This produces a diagram topologically equivalent to that shown in $d$ (here the blue, (yellow, mauve) lines represent representations in the tensor product of the red and green (black and green, red and black) Wilson line representations).  The diagram can be evaluated using the fusion rules of the appropriate quantum group. }
\end{center}
\end{figure}

\end{widetext}

For $T^2 \times S^1$ the prescription is exactly the same as for the torus, except that now there are three edges, and  two plaquettes which border every edge. 
To make the \ch link, we take a sphere with six punctures : top and bottom, to which we will attach the handle in the time-like direction; front and back, to which we will attach handle $a$ of the torus, and left and right, to which we attach handle $b$ of the torus.  The link can be deduced from the trajectories of the attaching maps of the $2$-handles on the sphere
 (Fig. \ref{Fig_T2S1CH}).  For example, the $2$-handle which represents the surface of the torus follows a trajectory $a b a^{-1} b^{-1}$, shown as the green numbered sequence in the Figure.  
 
 The final link $L_{T^3}$ is illustrated in Fig. \ref{Fig_T2S1CH}c.  Enforcing the constraint that the net flux through each of the yellow rings vanishes gives Fig. \ref{Fig_T2S1CH}d.  The invariant can be evaluated from this diagram to give
 \be
\mac{Z}_{T^2 \times S^1} = \frac{1}{\mac{D}^2} \langle L_{T^3} \rangle = r^2
\ee
where $r$ is the number of representations of the quantum group.

\subsection{Relation to lattice models}

The simple handle decompositions described above are convenient for constructing and evalulating the \ch link.  To relate these to lattice models, we modify these handle decompositions by adding extra $0$-handles (vertices), $1$-handles (edges), $2$-handles (plaquettes) and $3$-handles (to fill in any holes).  This deforms the simple \ch link to one which locally resembles Fig. \ref{Chain}, but does not affect the value of $\langle L_{\Omega} \rangle$\cite{RobertsThesis}.

\begin{figure}[h!]
\begin{center}
\includegraphics[width=0.8\columnwidth]{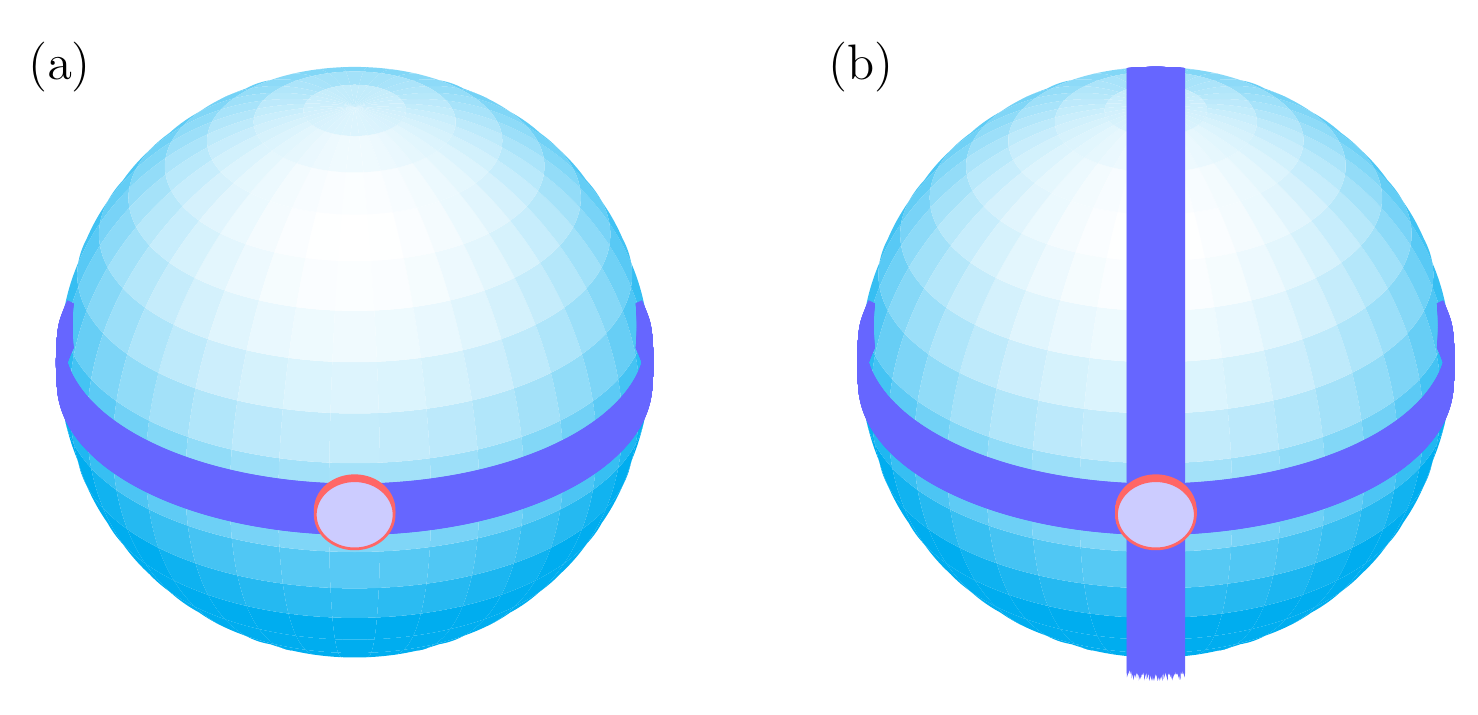}
\caption{ \label{Fig_S2S1_2}  (a) The spatial part of an alternative handle decomposition of $S^2 \times S^1$, where the sphere is divided into two plaquettes using a single $0$ handle (mauve), a $1$-handle (dark blue) and two $2$-handles (light blue).  (b) A lattice tiling on $S^2$.   To extend it to $S^2 \times S^1$, we simply add a $2$-handle between the two copies of each new $1$-handle on the sphere, and $3$-handles to fill in the new holes. }
\end{center}
\end{figure}

We demonstrate how this works for the space-time $S^2 \times S^1$.  To divide the sphere in two, we must add one $1$-handle (the edge between the two halves of the sphere), and one new $2$-handle to embody the new spatial plaquette.   We also add a new $2$-handle in the time-like direction, making a time-like plaquette over this edge (not shown in Fig. \ref{Fig_S2S1_2}).   Finally we must add an extra $3$-handle, since the time-like plaquette has divided the area between the initial and final time steps into two halves.   To make a more complicated lattice, we could add more $0$-handles (vertices), $1$-handles (edges), and $2$-handles (plaquettes) to the picture, as shown in Fig. \ref{Fig_S2S1_2}b.  This method can be used to create a trivalent lattice of arbitrarily many plaquettes, as desired for a Levin-Wen model -- though on the sphere not all plaquettes can be hexagons. 

\begin{figure}[h!]
\begin{center}
\includegraphics[width=0.9\columnwidth]{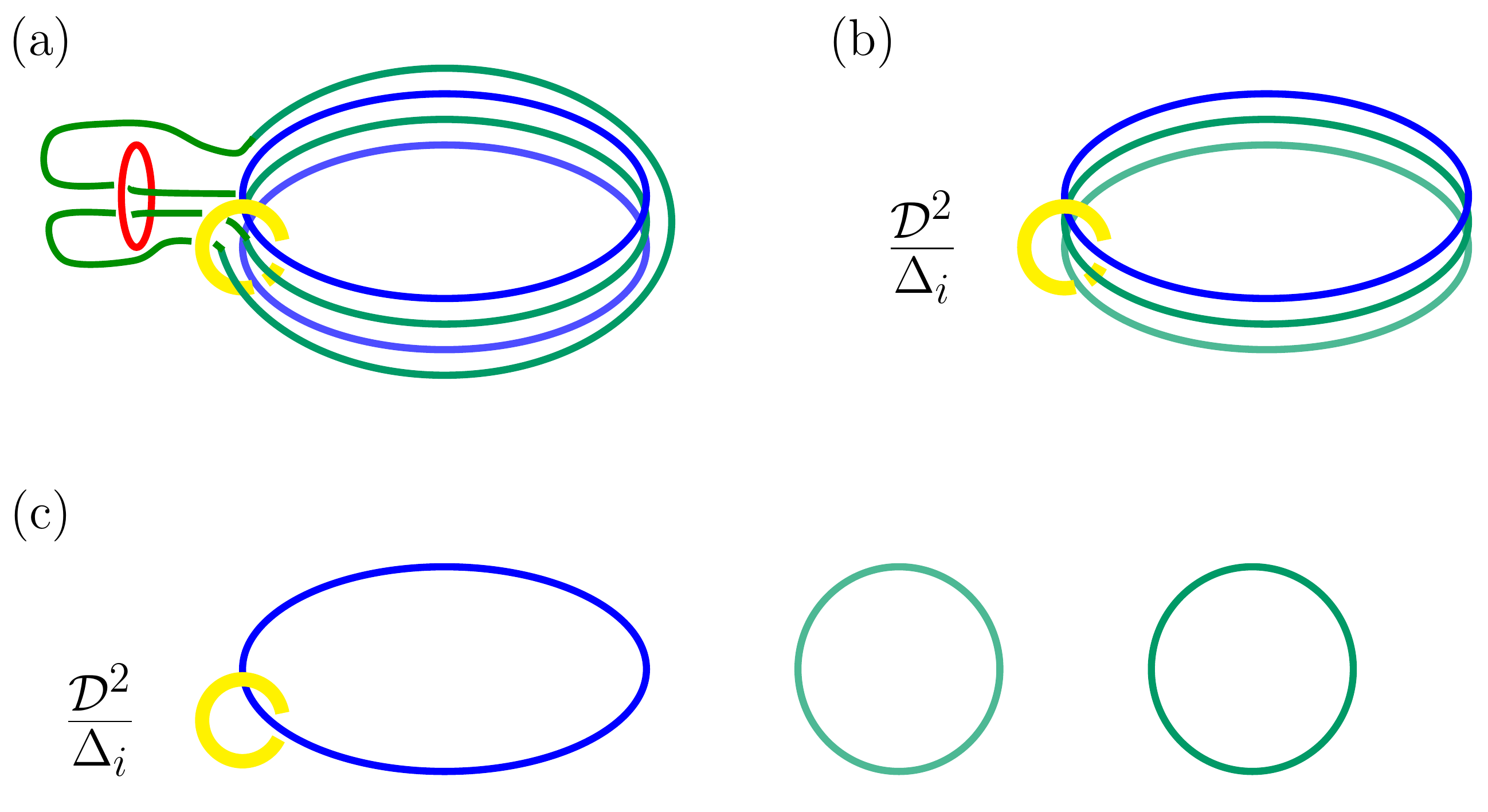}
\caption{ \label{S2Link2}  \ch link corresponding to the alternative handle decomposition of Fig. \ref{Fig_S2S1_2}a.  (a) The \ch link consists of three `plaquette' strings which trace the attaching maps of the two spatial 2-handles (blue) and the time-like 2-handle (green).  There are also two `edge' strings -- one for the one-handle extending around the sphere (yellow), and one for the one-handle extending in the time direction (red).  (b) The two blue strings trace the same curve, and can be combined to give a single loop (still labeled by $\Omega$) with an overall coefficient of $\mac{D}$.  The red loop's role is to ensure that the combination of the two green strings passing through its center carries no net flux; we eliminate it loop in favor of the diagram shown here with a coefficient $\mac{D}/\Delta_i$, where $i$ is the label carried by the green string.  (c) The green strings may handle-slide over the blue string, to give the diagram shown (with the coefficient $\mac{D}^2/\Delta_i$, from the eliminations performed in (b). )  This gives $\mac{Z}_{S^2 \times S^1} =1$. }
\end{center}
\end{figure}

The relevant \ch link is shown in Fig. \ref{S2Link2}a.  It consists of a pair of projector loops (blue) for the two spatial plaquettes, and a single time-like loop (green) carrying the index $i$ associated with the single edge in our lattice.  The loop around the time-like $1$-handle (red) can be identified with the vertex projector for the single vertex, while the loop around the spatial $1$-handle implements the action of the plaquette projectors on the edge label $i$.  

It is easy to see that evaluating $\langle L_\Omega \rangle$ for $L$ derived from the handle decomposition (Fig. \ref{Fig_S2S1_2}a) still gives the partition function on $S^2 \times S^1$ -- namely, unity.    A series of simple manipulations (described in the Figure caption) reduce this to the diagram shown in Fig \ref{S2Link2}c.  To evaluate the partition function from here, we note that :
% The $1$-handle in the time-like direction is crossed by two strings, since the temporal $2$-handle is glued onto it once going forwards, and once backwards.  We may use the fact that the loop enclosing these (red in the Figure) projects onto $0$ flux  to fuse these two strings together.   If the label of this string is $i$, then the result of this fusion is a coefficient of $\mac{D}/\Delta_i$, and a pair of strings labeled $i$ running around the edge of the space-like plaquette.   Now the diagram contains a projector string encircling the space-like edge, and four strings running along it.  Two of these came from the time-like plaquette, and carry the same label $i$.  The other two-- the two from the pair of spatial plaquettes -- are projectors, since they have not been affected by the fusion process.  We may combine these two to obtain a single projector string, along with a factor of $\mac{D}$.  Next, we handle-slide the pair of strings labeled $i$ over this projector loop, to obtain two unlinked unknotted $i$ strings, and a pair of linked projector strings. 
(1)  the linked $\Omega$ loops contribute a factor of unity, and (2) when the green string is labelled $i$, the pair of green loops contribute a factor of $\Delta_i^2$.  Since (from $\Omega = \frac{1}{\mac{D} } \sum_i \Delta_i i$) the label $i$ initially appeared with a coefficient $\Delta_i/ \mac{D}$, and including the factor $\frac{\mac{D}^2}{\Delta_i}$ from the manipulations in the Figure, we obtain $\mac{Z} = \mac{D}^{1 - n_0 - n_3}  \sum_i \Delta_i^2 = 1$.  
%Since there are two $3$-handles and one $0$-handle, this gives a value for the \ch link of unity, as expected.  

%More generally, no matter how we choose to slice up the surface of the sphere into a lattice, computing the \ch link will always give $\mac{Z} =1$, since the \ch link invariant does not depend on what handle decomposition we choose\cite{RobertsThesis}.  

\subsection{Evaluating the \ch link} \label{knotApp}

Her we list basic facts about the \ch link invariant which can be used to compute $\langle L_\Omega \rangle$.
The indices $i \in \{ 1 ... r \}$ of the lattice model correspond to labels which we can assign to the strands of the link.  
The notation $L_{\Omega}$ indicates that each stand of the link $L$ should labeled with the superposition $\Omega = \sum_i \Delta_i i$.

 To evaluate more involved link invariants requires {\it fusion} coefficients which we will not introduce here. The technical background for these calculations is given in Ref. \onlinecite{Us}.   However,  some simple link invariants can be easily evaluated using the following facts.  (1)  An $\Omega$ string that is unlinked from all other strings, and not twisted or knotted with itself, contributes a factor of $\mac{D}$.
(2) Every $\Omega$ string in the link acts like an un-normalized projector, which forces all of the strings inside of it to fuse together.  This process always incurs a factor of $\mac{D}$ (as the projector is not normalized), as well as some coefficients due to the fusion.   In particular, when a string $i$ is fused with its conjugate to give the identity, the coefficient for this process is $\mac{D}/\Delta_i$.
(3) $\Omega$ strings may `handle-slide' over each other, as shown in Fig. \ref{KillingHandle}, without changing the value of the link invariant.  
(4) When two $\Omega$ strings follow exactly the same trajectory, we may combine them into a single $\Omega$ string, incurring a factor of $\mac{D}$.

These rules are sufficient to evaluate simple link diagrams, such as those pertaining to $S^2 \times S^1$ above.  
More generally, there are a set of fusion rules dictating what coefficients are induced by this fusion, and how to change the order of fusion processes and thereby evaluate the diagrams.  In more involved examples, such as the link for $T^2 \times S^1$ above, it becomes necessary to use these to evaluate diagrams.

\end{appendix}

\acknowledgments

The authors acknowledge helpful discussions with Z. Wang, M. Freedman, K. Walker, and M. Levin.


\begin{thebibliography}{50}
\bibitem{channels} Most generally, the Hilbert space is supplimented by a fusion channel at each vertex when the neighboring edges can fuse to the identity in multiple ways.  See \cite{LevinWen}.
\bibitem{endnote1} Our construction will more generally apply to any lattice model whose spectrum is the double of a modular tensor category -- see Ref. \onlinecite{Us}.
\bibitem{handleslide} Handleslide is a more general manipulation of links, but we will only need this simple form of it (See Ref.~\cite{Wang})
\bibitem{endnote3} Strictly speaking one generates  ${\cal M} \# \overline {\cal M}$ with some additional (but unimportant) holes in it.  See \onlinecite{Martins,Us}.
\bibitem{NoteA} A useful introduction to handle decompositions is given in Ref. \onlinecite{Gompf}.
\end{thebibliography}
\end{document}